\documentclass[conference]{IEEEtran}
\usepackage{amsmath}
\usepackage{booktabs}
\usepackage{multirow}
\usepackage{tikz}
\usepackage{hhline}
\usepackage{cite}
\usepackage{graphicx}
\usepackage{pifont}
\usepackage{xcolor,colortbl}
\usepackage[parfill]{parskip}
\usepackage{amssymb}
\usepackage{hyperref}
\usepackage{cleveref}

\definecolor{LightCyan}{rgb}{0.84,1,1}
\definecolor{Gray}{gray}{0.85}
\newcolumntype{a}{>{\columncolor{LightCyan}}c}
\newcolumntype{b}{>{\columncolor{Gray}}c}

\newcommand{\makefigpdf}[4][0.5]{
\begin{figure}[t!]
    \centering
    \includegraphics[width=#1\textwidth]{images/#2.pdf}
    \caption{#3}
    \label{fig:#2}
\end{figure}
}

\newcommand{\cn}[1]{\tikz[baseline=(X.base)] 
  \node[draw,circle,fill=black,text=white,inner sep=0.8pt,font=\scriptsize\bfseries](X){#1};}

\usepackage{algpseudocode}
\usepackage{algorithm}
\usepackage{caption}
\usepackage{subcaption}
\usepackage{xcolor,soul}
\usepackage{xspace}
\usepackage{tabularx}

\newcommand{\figvs}[4]{\begin{figure}[!t]
\centering
\includegraphics[width=#1\columnwidth,keepaspectratio,#3]{images/#2}
\caption{#4}
\label{#2}
\end{figure}}

\newcommand{\comment}[1]{}

\newcommand{\base}{baseline\xspace}

\newcommand{\ddfive}{DD5\xspace}

\newcommand{\ddsix}{DD6\xspace}
\newcommand{\DD}{Double-Duty\xspace}

\IEEEoverridecommandlockouts
\IEEEaftertitletext{\vspace{-0.8\baselineskip}}

\begin{document}

\title{Double Duty: FPGA Architecture to Enable Concurrent LUT and Adder Chain Usage}

\author{
  \IEEEauthorblockN{Junius Pun\textsuperscript{*}}
  \IEEEauthorblockA{\textit{Nanyang Technological University} \\ 
                   JPUN001@e.ntu.edu.sg}
  \and
  \IEEEauthorblockN{Xilai Dai\textsuperscript{*}}
  \IEEEauthorblockA{\textit{Cornell University} \\
                   xd44@cornell.edu}
  \and
  \IEEEauthorblockN{Grace Zgheib}
  \IEEEauthorblockA{\textit{Altera} \\
                   grace.zgheib@altera.com}
  \and
  \IEEEauthorblockN{Mahesh A. Iyer}
  \IEEEauthorblockA{\textit{Altera} \\
                   mahesh.iyer@altera.com}
  \and
  \IEEEauthorblockN{Andrew Boutros}
  \IEEEauthorblockA{\textit{University of Waterloo} \\
                   andrew.boutros@uwaterloo.ca}
  \and
  \IEEEauthorblockN{Vaughn Betz}
  \IEEEauthorblockA{\textit{University of Toronto} \\
                   vaughn@ece.utoronto.ca}
  \and
  \IEEEauthorblockN{Mohamed S. Abdelfattah}
  \IEEEauthorblockA{\textit{Cornell University} \\
                   mohamed@cornell.edu}
}

\maketitle
\begingroup
\renewcommand\thefootnote{}\footnotetext{*~These authors contributed equally to this work.}
\endgroup

\begin{abstract}
Flexibility and customization are key strengths of Field-Programmable Gate Arrays (FPGAs) when compared to other computing devices.
For instance, FPGAs can efficiently implement arbitrary-precision arithmetic operations, and can perform aggressive synthesis optimizations to eliminate ineffectual operations.
Motivated by sparsity and mixed-precision in deep neural networks (DNNs), we investigate how to optimize the current logic block architecture to increase its arithmetic density.
We find that modern FPGA logic block architectures prevent the independent use of adder chains, and instead only allow adder chain inputs to be fed by look-up table (LUT) outputs.
This only allows one of the two primitives---either adders or LUTs---to be used independently in one logic element and prevents their concurrent use, hampering area optimizations.
In this work, we propose the Double Duty logic block architecture to enable the concurrent use of the adders and LUTs within a logic element.
Without adding expensive logic cluster inputs, we use 4 of the existing inputs to bypass the LUTs and connect directly to the adder chain inputs. 
We accurately model our changes at both the circuit and CAD levels using open-source FPGA development tools.
Our experimental evaluation on a Stratix-10-like architecture demonstrates area reductions of 21.6\% on adder-intensive circuits from the Kratos benchmarks, and 9.3\% and 8.2\% on the more general Koios and VTR benchmarks respectively.
These area improvements come without an impact to critical path delay, demonstrating that higher density is feasible on modern FPGA architectures by adding more flexibility in how the adder chain is used. 
Averaged across all circuits from our three evaluated benchmark set, our Double Duty FPGA architecture improves area-delay product by 9.7\%.

\end{abstract}


\section{Introduction}

Field-Programmable Gate Arrays (FPGAs) offer bit-level programmability at the price of performance when compared to Application-Specific Integrated Circuits (ASICs).
This enables their flexible use in many application domains where deployment-time reconfigurability is warranted such as cloud networking~\cite{firestone:nsdi18:azure,tarafdar:fpga17:network-fpga-clusters,:weerasinghe:fpt16:Network-attached}, wireless communication~\cite{havinga:fccm23:accelerating-Wi-Fi,wu:ieee2016:mu-mimo-ofdm,hassan:fccm24:mimo}, and machine learning~\cite{abdelfattah:fpl18:dla,zeng:fpga24:flightllm,wang:fpga22:shrinkage,wang:fpga24:grasu}. 
Many mainstream machine learning workloads depend on high-performance dense matrix multiplications, pushing FPGA vendors to either bolster their FPGAs with tensor compute units~\cite{langhammer:fpga21:stratix_nx,arora:fpga21:tensor_slices} or resort to heterogeneous FPGA-CGRA hybrid computing devices~\cite{gaide:fpga19:versal}.
However, as machine learning becomes ever more prevalent, it is finding its way into a broad variety of applications that can take advantage of the flexibility of the FPGAs' soft logic. 
This is highlighted by the plethora of works to create new frameworks~\cite{hls4ml}, methods~\cite{lutnet, wang:fpga22:shrinkage,polylut,ternary}, benchmarks~\cite{dai:fpl24:kratos}, and architectures~\cite{boutros:fpga19:math_hard,rasoulinezhad:fpga20:luxor} to improve the efficiency of deep neural networks (DNNs) implemented using the FPGA fabric soft logic.

Most DNN accelerators struggle to leverage unstructured sparsity or arbitrary mixed precision for higher efficiency~\cite{nvidia}.
This is also true for hard blocks (DSPs and BRAM) that are available on modern FPGAs.
For example, DSP units are often designed to optimize fused multiply-add operations that are 18, 27, or 32 bits wide~\cite{inteldsp, amddsp}. 
Despite many efforts to make FPGA hard blocks more reconfigurable~\cite{boutros:fpl18:diversity,arora:fccm22:comefa,chen:fccm23:bramac,chen:fpl23:m4bram}, they are still fundamentally limited by their modes of operation, their fundamental multiply-accumulate operation, and their limited resource number on a given FPGA chip.
In contrast, a soft-logic implementation of DNNs plays to the FPGA's strength of bit-level programmability by constructing exactly the circuit that is needed for an arithmetic operation.
This includes the flexibility to optimize away ineffectual operations at the operand and bit levels~\cite{wang:fpga22:shrinkage,dai:fpl24:kratos}, and to utilize custom numerical formats~\cite{fractal,carmichael:date19:positron}.
For instance, when synthesizing a multiplication operation where one operand is known at compile time, the operation can often be decomposed into a series of additions, allowing better utilization of FPGA adder chains.
Furthermore, adder trees are important and common as they are needed to perform reduction in matrix multiplication.

Several prior works have explored arithmetic efficiency in FPGA soft logic. 
Compressor trees, constructed with generalized parallel counters (GPCs) \cite{parandeh:asp-dac2008:comp-trees-lut, kumm:fpl2014:comp-trees-ilp, hossfeld:acm2024:comp-trees-amd}, combine the use of both LUTs and dedicated adders to further boost resource utilization. 
Other approaches involve architectural changes to the logic block, such as packing more adders in one logic element \cite{eldafrawy:acm2020:s10-mod} or including explicit XOR gates to improve compressor tree implementation~\cite{rasoulinezhad:fpga20:luxor}. 
However, current FPGA architectures present a fundamental limitation: the adder carry chain can only be driven by LUT outputs. 
Consequently, LUTs and adder chains must implement closely related logic to effectively utilize both resources. 
If a circuit is dominated by adder chains---as is often the case with matrix multiply reduction operations---the associated LUTs become unnecessarily occupied. 
Conversely, if LUTs implement unrelated logic functions, the adder chain remains inaccessible despite available inputs and outputs within the very same logic block.

To address this shortcoming, we propose a modification to FPGA logic blocks to more flexibly enable the concurrent use of adder chains and LUTs.
We fully implement our architecture---\DD{}---in open-source FPGA CAD tools, from circuit and architectural modeling, to synthesis algorithms, to enable an extensive evaluation over three popular FPGA benchmark suites.
Our results consistently show that the extra flexibility of enabling concurrent and independent use of LUTs and adder chains makes \DD{} significantly more area-efficient compared to current FPGAs without compromising critical path delay.
More concretely, we make the following contributions:
\begin{enumerate}
    \item We propose the \DD{} FPGA logic block architecture, which decouples the connections between LUTs and adders, enabling their independent and concurrent usage.
    \item We quantify the area and critical path delay of the additional circuit components introduced by the \DD{} architecture using SPICE-based simulation and transistor sizing with COFFE~\cite{chiasson2013coffe}.
    \item We integrate compressor tree algorithms into the open-source FPGA CAD tool Verilog-to-Routing (VTR)~\cite{vtr9}, significantly improving efficiency when synthesizing adder chains, ensuring a strong baseline on which to evaluate \DD{}.
    \item We evaluate our \DD{} architecture on a suite of comprehensive benchmarks, including the VTR standard benchmarks, Koios ML benchmarks~\cite{arora:ieee2023:koios}, and Kratos unrolled DNN benchmarks~\cite{dai:fpl24:kratos}, achieving an average 9.7\% improvement in area-delay product over all circuits, and up to 80\% increase in packing density in stress tests.
\end{enumerate}

\section{Background}

\subsection{FPGA Basics}

FPGAs can implement any logic function through programmable Lookup Tables (LUTs), organized into large clusters called Logic Blocks (LBs).
Every LB contains multiple (typically 10) smaller units, called Adaptive Logic Modules (ALMs) or Fracturable Logic Elements (FLEs), connected to a programmable interconnect network via a local crossbar. 
\Cref{baseline} shows the ALM design typical of Intel’s Stratix 10 series FPGAs. 
Each ALM typically consists of four 4-input LUTs that can be combined with multiplexers to implement two 5-input truth tables or a single 6-input truth table. 
In addition to LUTs, each ALM contains two 1-bit full adders, whose inputs are directly connected to the output of 4-LUTs. 
This structure helps simplify logic before addition, and the LUT can be used in many addition algorithms \cite{hossfeld:acm2024:comp-trees-amd,parandeh:asp-dac2008:comp-trees-lut}. 
Moreover, the carry-in and carry-out signals are connected along multiple ALMs, forming a long carry chain, allowing fast and high-bit-width integer addition.

Despite this versatility, a key limitation of this conventional FPGA architecture is that the LUTs and adders are not independent. 
When the adders are in use, the LUT outputs provide inputs to the adder chain and thus cannot be used to implement other logic functions. 
This dependence limits the resource utilization in many arithmetic-heavy workloads. 
Some of the recent commercial FPGAs from Xilinx \cite{amdlut} have completely removed the 1-bit adders and instead use LUTs to generate carry propagate and generate signals, which further limits the ability of using arithmetic and logic resources independently.

\subsection{CAD Tools}

\textbf{VTR} Verilog-to-Routing (VTR)~\cite{vtr8, vtr9} is an open-source FPGA CAD tool that takes Verilog design files with an FPGA architecture description file and performs synthesis, placement, routing, and timing analysis. By using a tree-structured XML architecture description file, VTR allows users to experiment with arbitrary FPGA architectures and quantitatively evaluate new architectures and CAD algorithms.

\textbf{Parmys} VTR has undergone several enhancements over time. Recently it changed the synthesis front-end from Odin II~\cite{odinii} to Yosys~\cite{yosys}, a more flexible and advanced RTL synthesis tool that supports modern features like SystemVerilog generate statements. 
However, as the Yosys front-end is a general-purpose synthesis tool, VTR relies on a plugin called Parmys (Partial Mapper for Yosys) to handle FPGA-specific technology mapping~\cite{vtr9}. 
Parmys has some significant limitations, particularly in the optimization of arithmetic operations including addition and multiplication. 
In this paper, we develop adder chain synthesis within Parmys to significantly improve its efficiency in synthesizing adder chains and compressor trees, to be able to investigate our proposed \DD{} architecture using a strong CAD baseline.

\textbf{COFFE 2} Circuit Optimization For FPGA Exploration (COFFE) \cite{chiasson2013coffe} is an automated transistor-level modeling tool that provides accurate estimates of area, delay, and energy consumption for FPGA tiles, by utilizing HSPICE simulations and automated transistor sizing. Its successor, COFFE 2\cite{coffe2}, extends these capabilities to heterogeneous FPGA architectures, supporting complex logic blocks, fracturable LUTs, and custom DSP tiles. In this work, we use COFFE 2 to model our \DD{} architecture for precise area and timing estimation.

\subsection{Arithmetic Optimizations}

\begin{figure}[t]
    \centering
    \includegraphics[width=0.8\columnwidth]{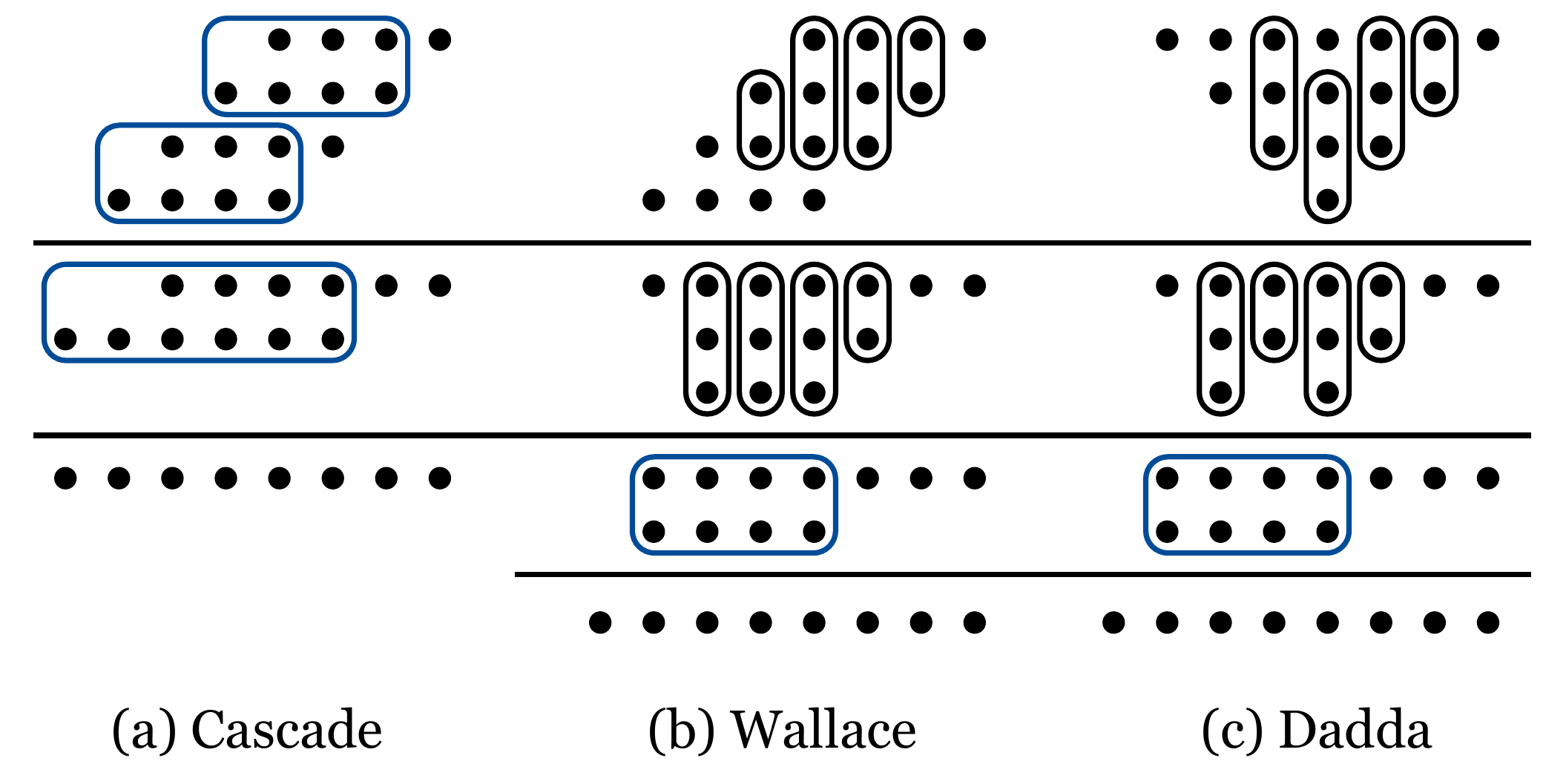}
    \caption{Conceptual diagram of a 4-bit multiplier using Cascade, Wallace, and Dadda algorithms. Pill shapes represent LUT-based compressors and rectangles represent full adder chains. Each layer represents a reduction stage.}
    \label{fig:trees}
\end{figure}

Integer multiplications are the core of computation workloads, especially in machine learning. 
While DSPs are specialized units dedicated to multiplication, they are a limited resource and are usually better utilized with high bit width, and when both operands are unknown at compile time. 
However, a soft logic multiplier is still crucial to optimize resource utilization, particularly for custom and low-bitwidth arithmetic, and especially for unrolled DNNs~\cite{lutnet, ternary, dai:fpl24:kratos} in which one of the operands is known at compile time---the DNN model parameters. 
This reduces a multiplication operation into multiple additions of partial products.
For an n-bit by n-bit multiplication, n such rows are formed, which require n-1 summations. 
Using only full adders, an approach to reduce latency is to use binary adder trees, which perform as many summations as possible in parallel, requiring $O(\log n)$ time to obtain the final result. 
This is currently the default and only approach taken by VTR to synthesize soft multipliers.
To exploit the available LUTs in FPGAs, compressor trees~\cite{kumm:fpl2014:comp-trees-ilp,hossfeld:acm2024:comp-trees-amd,parandeh:asp-dac2008:comp-trees-lut} can also be used for efficient summation.
Carry save logic, implemented with LUTs, is used to compress the initial n rows into 2 final rows that can be summed together with a fast ripple carry chain. \Cref{fig:trees} illustrates the conceptual diagrams of two widely used algorithms, Wallace and Dadda, with Cascade---a simple algorithm that only uses adder chains---for comparison.
Notably, compressor trees are currently not supported in Parmys/VTR.  

\section{\DD{} Architecture and Circuit-Level Modeling}

This section presents our new architecture, \DD{}, that enables the concurrent use of logic block LUTs and adder chains, resulting in denser FPGA placement.
We describe the architectural enhancements and variants, evaluate our proposal at the circuit level, and present the CAD synthesis changes that were needed to effectively and fairly assess our proposed architecture.

\subsection{\DD Logic Block Architecture}

\begin{figure}[t]
     \centering
     \begin{subfigure}[h]{0.47\textwidth}
         \centering
         \vspace{-0.2cm}
         \includegraphics[width=\textwidth, trim= 0 0 0 0]
         {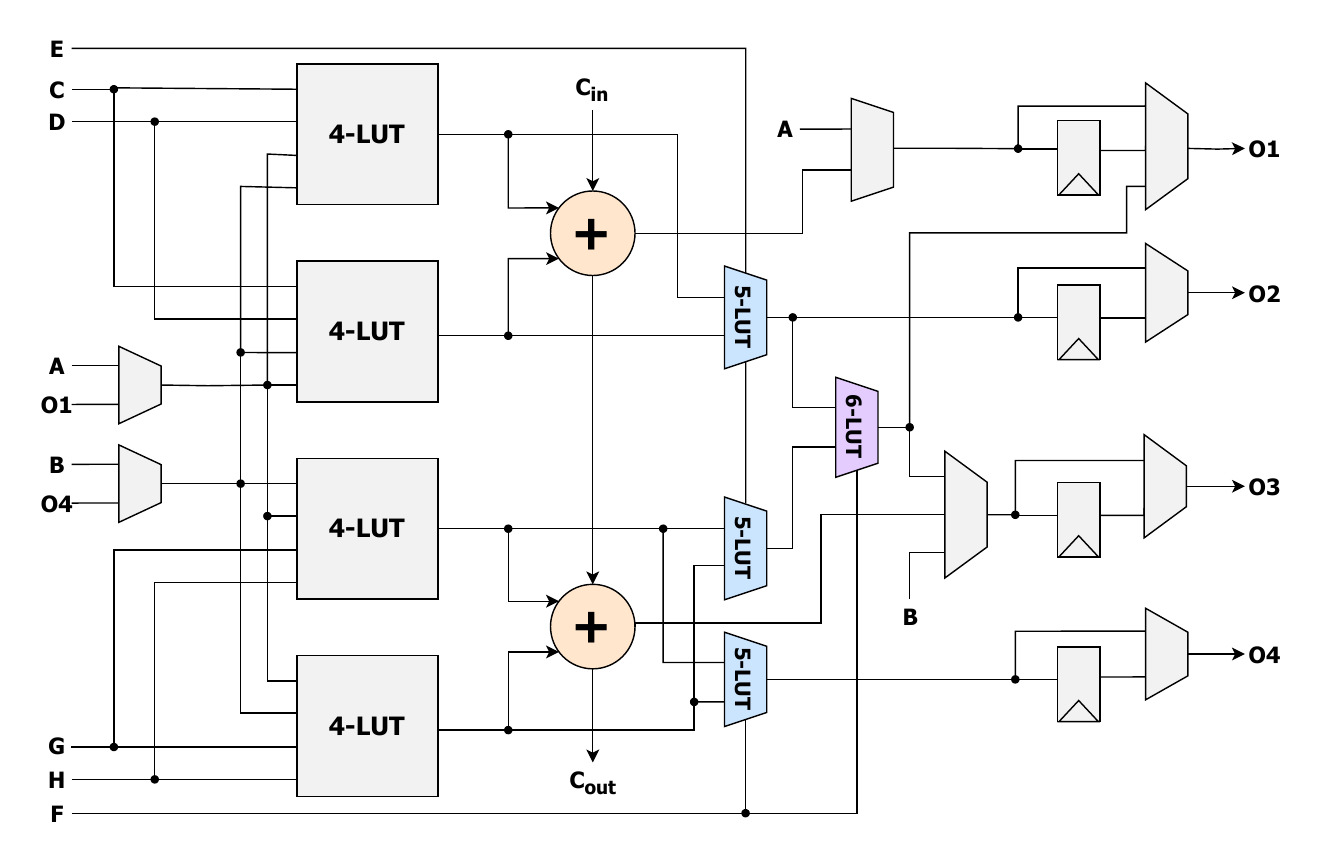}
         \caption{Baseline Stratix-10 ALM.}
         \vspace{0.5cm}
         \label{baseline}
     \end{subfigure}
     \vfill
     \begin{subfigure}[h]{0.47\textwidth}
         \centering
         \vspace{-0.2cm}
         \includegraphics[width=\textwidth, trim= 0 0 0 0]{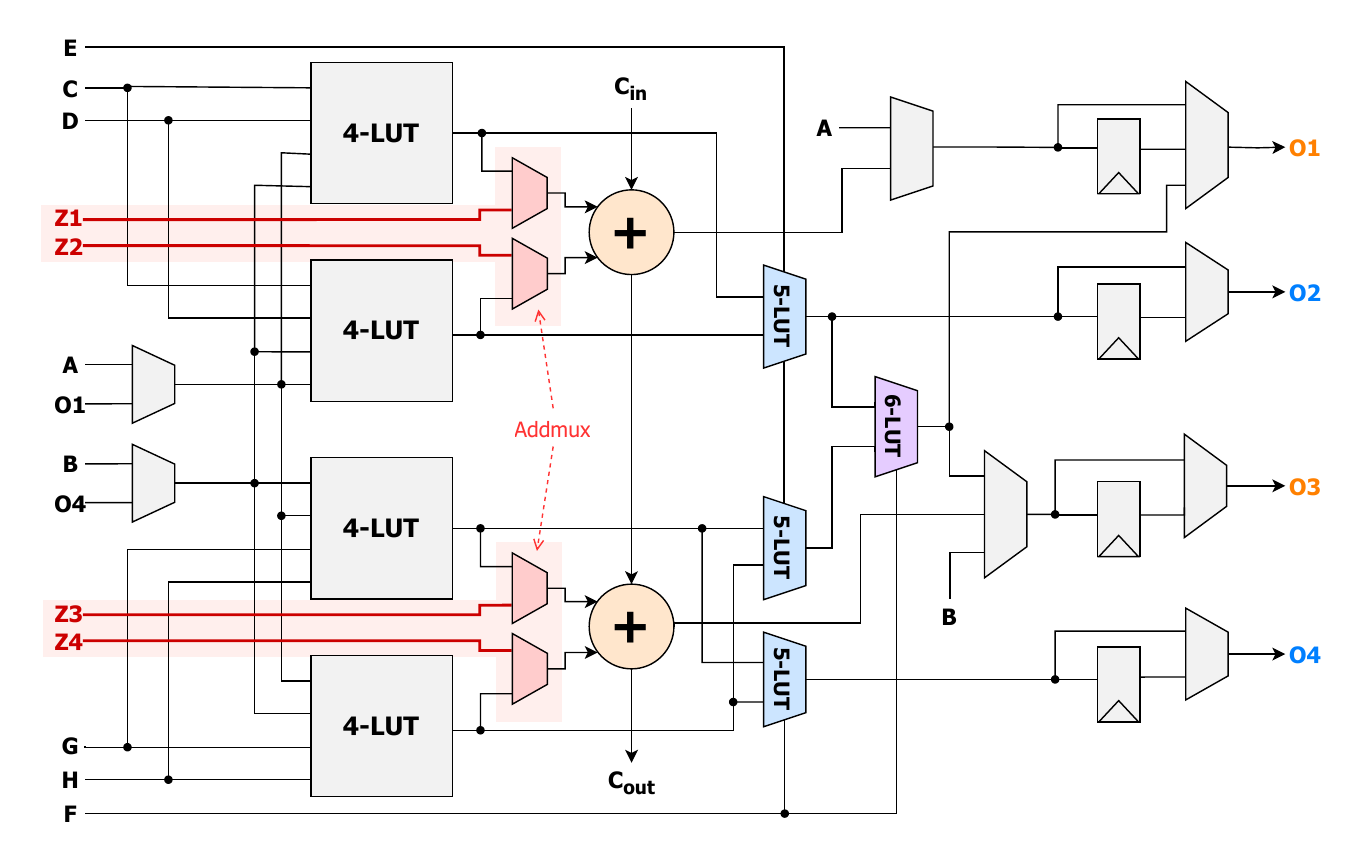}
         \caption{\DD ALM. (\ddfive)}
         \vspace{0.5cm}
         \label{dd_5lut}
     \end{subfigure}
     \vfill
     \begin{subfigure}[h]{0.47\textwidth}
         \centering
         \vspace{-0.2cm}
         \includegraphics[width=\textwidth, trim= 0 0 0 0]{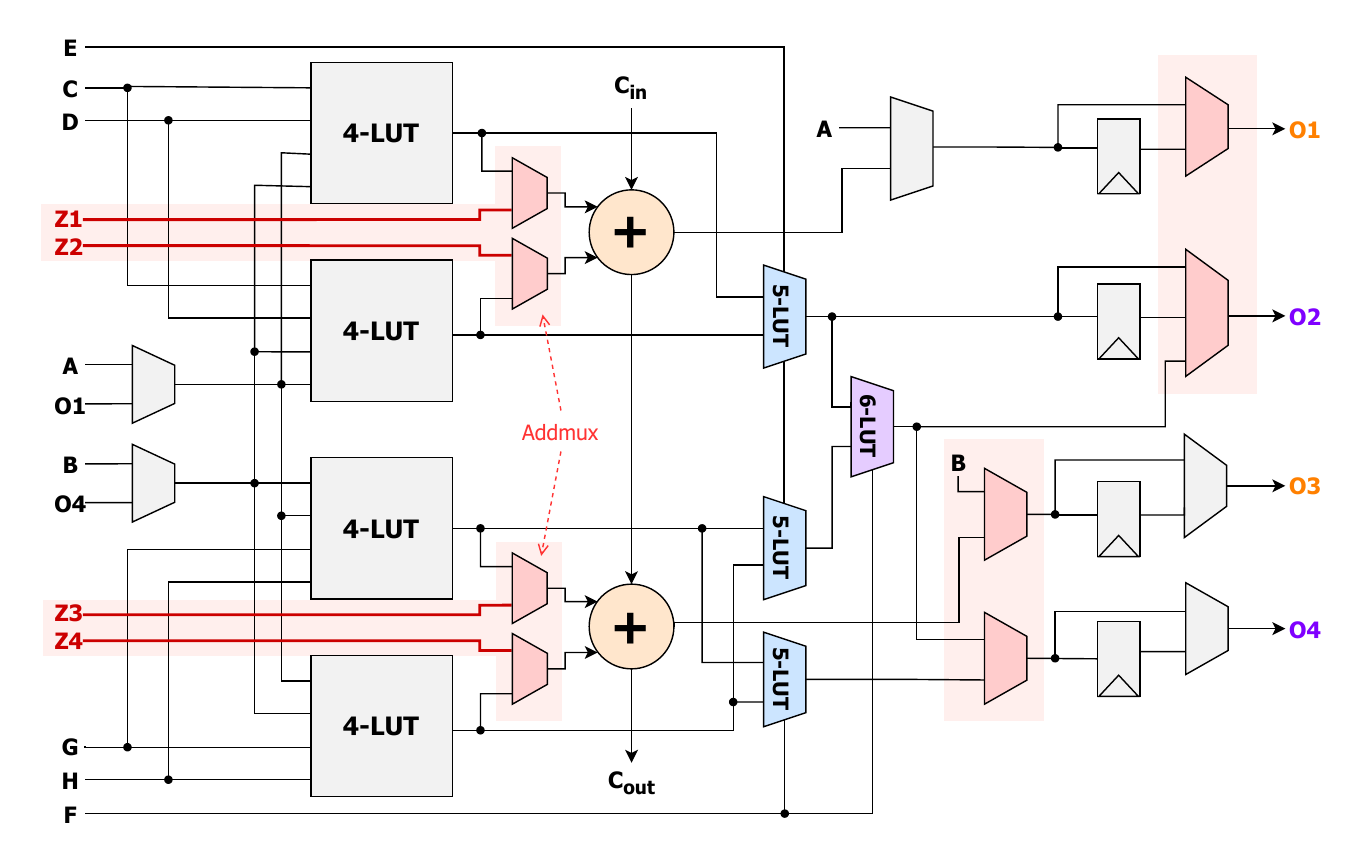}
         \caption{\DD ALM with concurrent LUT6 mode. (\ddsix)}
         \vspace{0.3cm}
         \label{dd_6lut}
     \end{subfigure}
     \vspace{6pt}
    \caption{Baseline (a) and \DD ALM architectures. We show two variants: the \ddfive (b) supports concurrent 5-LUT and adders usage while the \ddsix (c) supports concurrent 6-LUT and adder usage.}
    \label{fig:architecture}
\end{figure}

To enable independent and concurrent use of both LUTs and adders, we introduce two main changes to the ALMs and local routing as described below.
Furthermore, we present two variants \textbf{\ddfive{}} and \textbf{\ddsix{}} which enable the use of 5-LUT and 6-LUT modes concurrently with the adders, respectively.

\figvs{0.9}{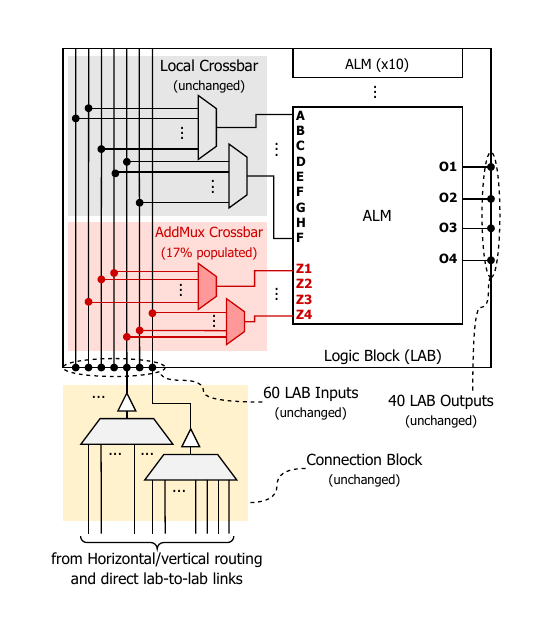}{trim = 0 20 0 20, clip}{A new AddMux Crossbar connects existing LB inputs to the new adder chain direct paths (Z1--Z4).}

\textbf{AddMux} In each ALM, we add extra multiplexers and four additional ALM inputs (Z1--Z4), as highlighted in \Cref{dd_5lut}. 
These modifications allow inputs to bypass the LUTs and connect directly to the adder chain, enabling the use of adders and LUTs concurrently and independently. 
In this revised architecture, the ALMs support independent operation in 5-LUT mode: two output pins are allocated for adder outputs (O1 and O3), while the remaining two output pins can be used for the 5-LUT outputs (O2 and O4). 

\textbf{AddMux Crossbar} To accommodate the additional ALM inputs (Z1--Z4), we introduce a secondary local interconnect, which is sparsely populated and draws inputs from the existing LB inputs as shown in \Cref{crossbar}. 
Importantly, the total number of input pins to the LB remains unchanged---this modification leverages the same inputs of the original LB, and therefore does not require any modifications to global routing. 
Notably, the wires that our new crossbar connects to have connections from LB-to-LB direct links~\cite{khoozani:fpl23:titan_2.0}.
Specifically, Stratix-10 architectures include 40 LB-to-LB wires, 20 from the east LB and 20 from the west LB. 
These enable direct connections between adjacent LBs without needing to access the global interconnect.
As shown in \Cref{crossbar}, these direct LB-to-LB connections are multiplexed together with global routing into the connection block multiplexers.
These feed 60 LB inputs, 40 of which can be connected to LB-to-LB inputs.
Our new AddMux Crossbar connects to 10 out of those 40 LB inputs, making this new crossbar only 17\% populated ($\frac{10}{60}$ inputs).
In contrast, the existing local crossbar is more than 50\% populated~\cite{khoozani:fpl23:titan_2.0}.

\textbf{DD6} The AddMux and AddMux Crossbar allow the concurrent usage of 5-LUTs and adders; however, \ddfive does not support concurrent 6-LUT usage with the adders. 
To address this, we propose an extended version, \ddsix, which modifies the multiplexing of LUT and adder signals to the outputs, enabling fully independent 6-LUT and adder operation, as shown in \Cref{dd_6lut}. 
However, our preliminary analysis indicates that the additional flexibility provided by \ddsix yields marginal benefits in practice. 
A detailed evaluation of these trade-offs is presented in Section~\ref{sec:eval}.

\subsection{Circuit-Level Modeling of \DD{}}

To model the delay and area overheads introduced by our architectural modifications, we build upon the framework established by Eldafrawy et al.\cite{eldafrawy:acm2020:s10-mod}, leveraging COFFE 2\cite{coffe2} to replicate the Stratix-10-like baseline architecture used in their work. 
In particular, we create SPICE models for AddMux, as well as the additional AddMux Crossbar. 
COFFE 2 is then used to automatically size the transistors, and the resultant area and delay characteristics are used in the VTR model for the benchmark-level evaluations. 
\Cref{tab:dd_added_components} shows the area and delay costs of each added component. 
Overall, our modifications increase the tile area by 3.72\% as compared to the baseline architecture. 
\Cref{tab:dd_delay_impact} shows the delay impact of the added components on several paths within the architecture.  
Although these multiplexers cause a LUT-to-adder delay increase, the delay of any signal feeding the adders directly is almost cut in half, as now the signal does not need to go through the LUT anymore. 
In addition, the delay of LUT output to ALM output remains almost the same as the output circuitry is largely unchanged.
Note that Table~\ref{tab:dd_delay_impact} shows that the AddMux Crossbar has higher delay than the existing Local Crossbar. 
This may seem counterintuitive, as the AddMux Crossbar is much smaller; however, this additional delay is an artifact of transistor sizing in COFFE 2. 
COFFE 2 tries to optimize all the paths going through an ALM, and the Z ALM inputs have a much lower delay to the output of the ALM compared to the conventional LUT inputs; therefore, COFFE 2 aggressively optimizes the Local Crossbar for delay but the AddMux Crossbar can afford to be much slower and thus use smaller transistors.
 
\begin{table}[t!]
    \footnotesize
    \centering
    \setlength{\tabcolsep}{3pt}
    \renewcommand{\arraystretch}{1.05}
    \caption{Area and delay of added circuit components. The area data are all shown per ALM.}
    \vspace{3pt}
    \begin{tabular}{lcc}
    \toprule
    \textbf{Circuit}        & \textbf{Area (MWTAs)}        & \textbf{Delay (ps)}    \\
    \midrule
        AddMux              & { 1.698}                     & {68.77}               \\
        Baseline Crossbar   & { 289.6}                     & {72.61}               \\
        AddMux Crossbar     & { 77.91}                     & {77.05}               \\
        \midrule
        Baseline ALM        & { 2,167.3}                    & { -}                   \\
        \ddfive ALM         & { 2,366.6 (\textbf{+3.72\%})} & { -}                   \\
    \bottomrule
    \end{tabular}
    \label{tab:dd_added_components}
    \vspace{5pt}
\end{table}
%
\begin{table}[t!]
    \footnotesize
    \centering
    \setlength{\tabcolsep}{3pt}
    \caption{Delay impact of added circuits in the \DD architecture variants on data paths.}
    \vspace{3pt}
    \resizebox{0.5\textwidth}{!}{
    \begin{tabular}{clc}
    \toprule
    \textbf{Architecture} & \textbf{Path} & \textbf{Delay (ps)} \\
    \midrule
        \multirow{2}{*}{Baseline} & {LB input $\rightarrow$ ALM inputs A-H \cn{1}} & { 72.61} \\
                       & {ALM inputs A-H $\rightarrow$ Adder input \cn{2}} & {133.4} \\
    \midrule
                       & {LB input $\rightarrow$ ALM inputs $Z_1$-$Z_4$} & { 77.05 (+6.11\% vs. \cn{1})} \\
        \DD            & {ALM inputs A-H $\rightarrow$ Adder input} & { 202.2 (+51.6\% vs. \cn{2})} \\
                       & {ALM inputs $Z_1$-$Z_4$ $\rightarrow$ Adder input} & { 68.77 (-48.4\% vs. \cn{2})} \\
    \bottomrule
    \end{tabular}
    }
    \label{tab:dd_delay_impact}
    \vspace{5pt}
\end{table}


\section{VTR CAD Enhancements}

Soft multiplication synthesis in VTR hasn't been fully optimized, which results in redundant and/or wasteful resource usage. 
To tackle these shortcomings, we integrated several optimization techniques into VTR's synthesis. This ensured a strong CAD baseline in the evaluation of our Double-Duty architecture. 

\textbf{Unrolled Multiplication} 
General multiplication treats each partial product as a unique signal, since its value is unknown during compilation. 
However, in the case of unrolled multiplication, the rows are shifted duplicates of the multiplicand, and are included in the summation if the corresponding bit of the known operand is ‘1’. 
We define this corresponding bit as the “selector bit” of the row. 
When forming partial sums from these rows, there can therefore be adder chains that have identical input signals but sum different rows. 
Instead of synthesizing duplicate adder chains, which is the current VTR behaviour, a single adder chain can be used instead, with its output signals fanned out as inputs for future adder chains. 
Multiplicand rows can also be excluded if their selector bits are ‘0’, reducing the initial number of rows to sum.
For a sample 8-bit multiplication with a $(01010101)_2$ constant, baseline VTR uses $2.85\times$ more full adders than in the optimal case which exploits adder chain redundancy.

\textbf{Improved Binary Adder Tree Synthesis}
For $n$ rows to sum, it is possible to insert $\lfloor n/2 \rfloor$ parallel adder chains at stage 1 of the addition operation, whose $k$ rows of outputs form the next $k$ rows to reduce in stage 2. 
Adder chains need not necessarily be formed from adjacent rows, which presents many possible adder chain combinations. 
To find the best combination at every stage, a \textit{strength heuristic} is defined to reduce the ratio of included signals by the adder chains to the number of output signals generated in this stage. 
Included signals are counted by position, meaning that a duplicate input signal in different rows can be counted multiple times. 
The heuristic thus rewards adder chain placements with duplicate adder chains that can be replaced with a single adder chain. \Cref{fig:parmys-strength} demonstrates this concept.
To determine the optimal placement, a dynamic programming approach (\Cref{algo:row_selection}) is used to find a placement that gives maximum strength for each stage. 

\makefigpdf[0.5]{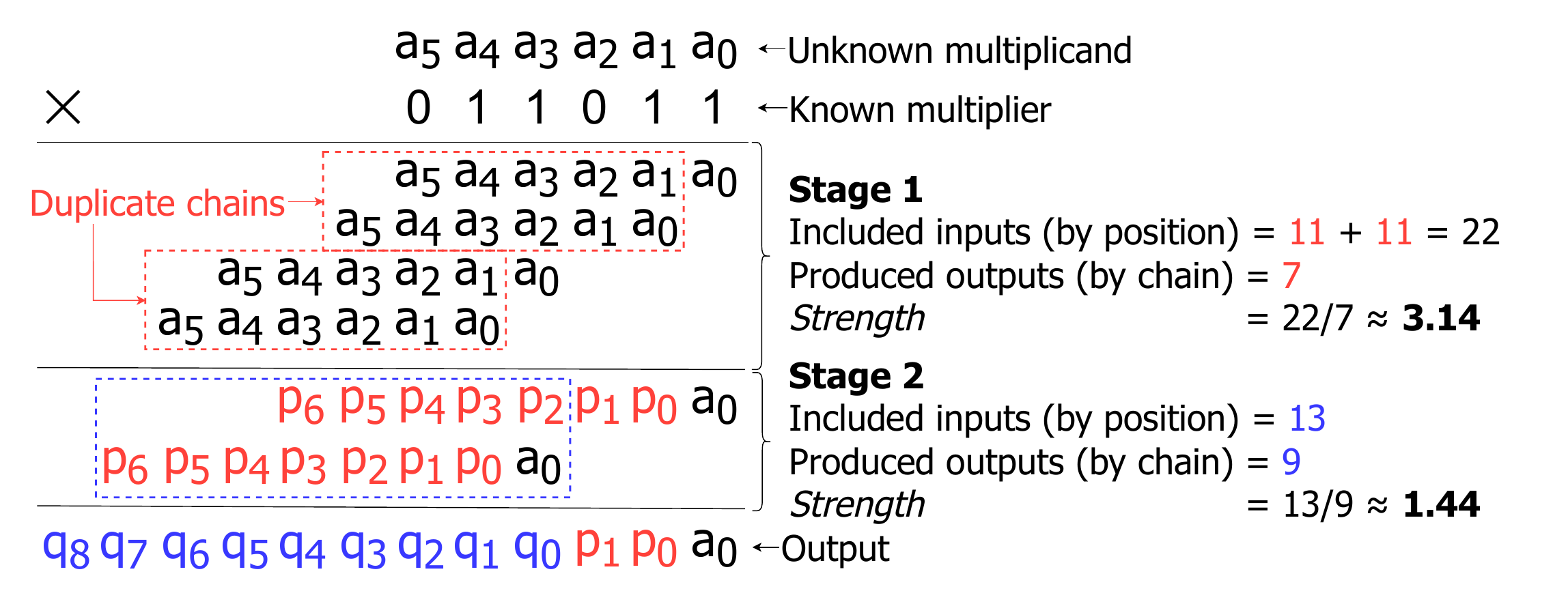}{A sample 6-bit by 6-bit unrolled multiplication. With known multipliers, it is possible to have duplicate adder chains in reduction stages. The \textit{strength} of each reduction stage is computed with input signals unique \textit{by position}, and output signals unique \textit{by chain}.}

\begin{algorithm} [b!]
    \footnotesize
    \caption{Adder row selection for maximum strength.}
    \label{algo:row_selection}
    \begin{algorithmic}
        \State $C_S \gets $ empty collection \Comment{empty solution cache}
        \Procedure{BestPlacement}{$R_n$} \Comment{$n$-subset of rows}
            \If{$n = 2$} \Return $R_n$ \EndIf
            \If{$C_S(R_n)$ exists} \Return $C_S(R_n)$ \EndIf

            \State $S_{\text{best}} \gets $ null \Comment{best solution}
            \If{$n$ is even}
                \For{$p \gets$ all pairs in $R_n$}
                    \State $S_{R_{n-2}} \gets$ \Call{BestPlacement}{$R_{n-2} \gets R_n \backslash \{p\}$}
                    \State $A_p \gets$ adder chain with $p$ as inputs 
                    \State $I_p, O_p, I_S, O_S \gets$ input, output signals of $A_p$, $S_{R_{n-2}}$ 
                    \State $I_S \gets I_S + I_p$
                    \If{$A_p$ is not used in $S_{R_{n-2}}$} $O_S \gets O_S + O_p$
                    \EndIf 
                    \If{$H_S = I_S / O_S > H_{S_{\text{best}}}$ or $S_{\text{best}}$ is null}
                        \State $S_{\text{best}} \gets S\{A_p,S_{R_{n-2}}\}$
                    \EndIf
                \EndFor
            \ElsIf{$n$ is odd}
                \For{$r \gets$ $R_n$} \Comment{for each row}
                \State $S_{R_{n-1}} \gets$ \Call{BestPlacement}{$R_{n-1} \gets R_n \backslash \{r\}$}
                \If{$H_{S_{R_{n-1}}} > H_{S_{\text{best}}}$ or $S_{\text{best}}$ is null} $S_{\text{best}} \gets S_{R_{n-1}}$
                \EndIf
                \EndFor
            \EndIf

            \State Add $S_{\text{best}}$ to $C_S$, \Return $S_{\text{best}}$
        \EndProcedure
    \end{algorithmic}
\end{algorithm}

\textbf{Compressor Tree Synthesis}
Instead of relying solely on adders, compressor trees implement reduction operations using both LUTs and adder chains. 
Current compressor tree implementations on FPGAs rely on generalized parallel counters (GPCs) specific to each architecture~\cite{hossfeld:acm2024:comp-trees-amd}. 
To implement compressor trees for any arbitrary architecture, each output signal of a compressor can be viewed as the result of a boolean equation of the GPC's input signals. 
After inserting compressors at each reduction stage, the final two rows are summed together with an adder chain. 
The intermediate combinational logic can then be optimized as part of logic synthesis, and then packed into LUTs. 
ABC synthesis~\cite{abc} is responsible for this logic packing within the VTR suite, and thus only the initial boolean logic mapping from the soft multiplier’s input signals to the final adder chain’s input signals needs to be implemented in VTR.
To achieve this, we make use of two compressor tree algorithms, the Proposed-Wallace (PW) and Dadda trees \cite{asif:vlsi2014:wallace-dadda}. 
PW and Dadda have been empirically shown to minimize and maximize the number of full adders required in the final adder chain, respectively. 
Both of these trees use only full and half adders as compressors, for which the boolean relationships between their inputs and outputs are well-known. 
Whenever the algorithm requires the use of a compressor, it is inserted as its logically equivalent combinational logic, constructed with boolean gates. 
The final compression logic is thus a combinational circuit that can be optimized as a whole and then synthesized into LUTs by ABC.

\begin{figure}[t]
    \centering
    \includegraphics[width=\columnwidth]{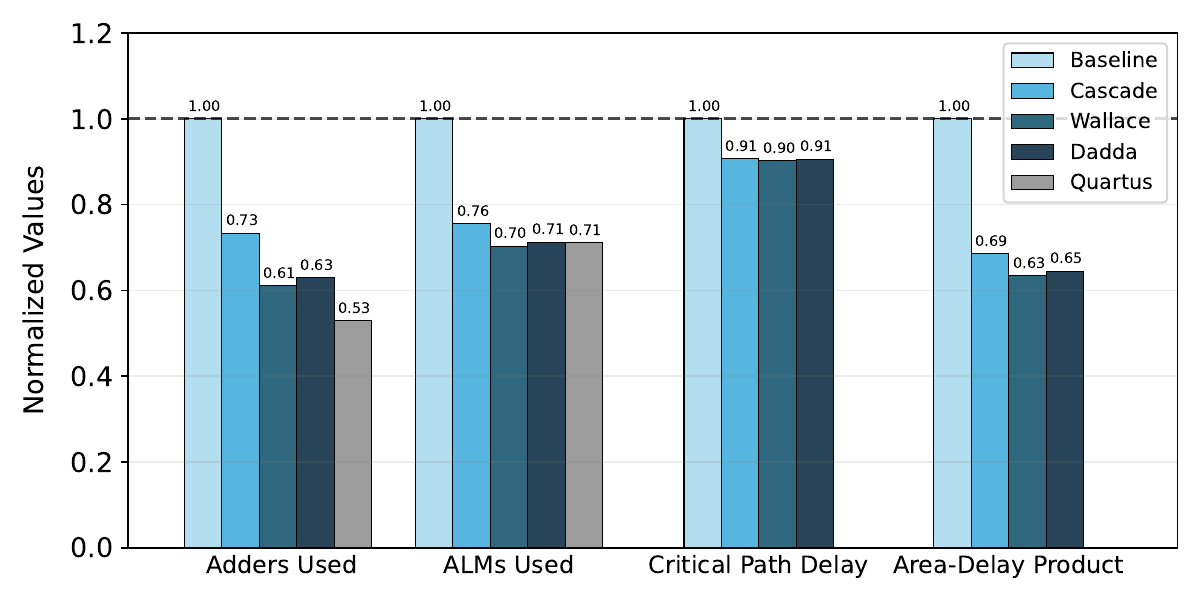}
    \caption{Normalized adders, ALMs, critical path delay, and area-delay product (ADP) comparison between baseline VTR and our improved Cascade, Wallace, and Dadda adder synthesis algorithms on the Kratos benchmark set. Quartus area results are also displayed for comparison.}
    \label{fig:VTR_validation}
\end{figure}

\textbf{CAD Improvement Validation}
After implementing the proposed synthesis algorithms and integrating them into VTR, we validated the changes using the Kratos benchmark suite. 
\Cref{fig:VTR_validation} summarizes the geometric mean of adder usage, ALM usage, critical path delay, and area-delay product across various benchmarks and operand widths on the Stratix 10 FPGA.
For reference, we also include the results of adder and ALM utilization from Intel Quartus Prime Pro 23.1 to compare resource efficiency with a commercial-grade tool. 
The goal is not to outperform Quartus which uses more advanced optimizations that we have not implemented. Instead, it is to validate whether our CAD flow generates realistic FPGA resource utilization and is representative of a commercial FPGA CAD flow.
We omit Quartus timing results as it uses a different timing model, and the open-source Stratix-10-like architecture in VTR does not provide accurate timing data. 
Overall, our updates to VTR improve upon the baseline by $\sim$37\% in terms of area-delay product, and closely match Quartus in terms of resource utilization. 
This provides a solid foundation for fair and realistic comparisons in our end-to-end evaluations presented in the next sections.

\section{Evaluation}
\label{sec:eval}

\subsection{Experimental Setup}

\textbf{VTR Setup}
The baseline FPGA architecture used for evaluation is modeled after Intel’s Stratix 10 architecture, with a 20-nm technology node for delay and area estimation. 
This model is directly available from the Verilog-to-Routing (VTR) repository and was developed in prior work\cite{eldafrawy:acm2020:s10-mod}.
We build upon this Stratix-10-like architecture by modeling the AddMux, AddMux Crossbar, and different output multiplexing to get the \ddfive and \ddsix architectures. 
Timing models for these components are obtained from COFFE 2.
The number of ALMs per LB is fixed at 10, as in the original architecture, and all other FPGA components remain unchanged.
We fix the channel width at 400, and we set the VTR \verb|target_ext_pin_util| to 0.9 for both inputs and outputs---this allows the VTR packer to use up to 90\% of the inputs and outputs of a logic block. 
The adder synthesis algorithm is set to use ``Wallace'' as this gives the best results in all metrics, as shown in \Cref{fig:VTR_validation}. 
The placement and routing are all timing-driven, and we run every experiment three times with three different seeds and take the average to get representative results.

\begin{table}[t]
\centering
\caption{Key statistics of our three benchmark suites, measured on the baseline Stratix 10 architecture with VTR.}
\resizebox{\columnwidth}{!}{
\begin{tabular}{l c cc cc c}
\toprule
\multirow{2}{*}[-0.27em]{\shortstack{Benchmark}}              & \multirow{2}{*}[-0.27em]{\shortstack{Num. \\Circuits}} & \multicolumn{2}{c}{ALMs ($\times 10^3$)}  & \multicolumn{2}{c}{Adder Percent}    & \multirow{2}{*}[-0.27em]{\shortstack{Avg. Fmax \\ (MHz)}} \\
                                                                                                   \cmidrule(l){3-4}                                  \cmidrule(l){5-6}
                                        &                                                          & Avg.            & Max                      & Avg.              & Max                 &                  \\
\midrule
VTR~\cite{rose:acm2012:vtr-standard-bm} &     19                                                   & 10.2               & 59.7                     & 19.5\%               & 47.7\%              &  109.5           \\
Koios~\cite{arora:ieee2023:koios}       &     20                                                   & 64.3               & 360.7                    & 22.5\%               & 50.2\%              &  70.9            \\
Kratos~\cite{dai:fpl24:kratos}          &      7                                                   & 59.6               & 208.2                    & 61.4\%               & 70.7\%              &  103.7            \\        
\bottomrule
\end{tabular}
}
\label{tab:benchmark_stats}
\end{table}

\textbf{Benchmark Selection}
To assess the impact of our architectural modifications, we evaluate our designs across the three benchmark suites shown below. Table~\ref{tab:benchmark_stats} also summarizes key information about each benchmark.

%
\begin{itemize}
    \item Kratos: A specialized set of circuits designed for unrolled deep neural networks (DNNs) \cite{dai:fpl24:kratos}. Kratos allows customized circuit sizes and varying computation sparsity levels, making it ideal for studying the effects of our optimizations. In this evaluation    , we used the same size of the ``small-size" set of benchmarks as in the original Kratos paper.
    \item Koios: A benchmark suite focusing on machine learning designs implemented on FPGAs \cite{arora:ieee2023:koios}. This benchmark set helps to assess how well our architecture adapts to typical ML workloads beyond Kratos.
    \item VTR Standard: A general-purpose FPGA benchmark suite from the VTR repository \cite{rose:acm2012:vtr-standard-bm}. This benchmark set ensures that our modifications do not degrade the versatility of the FPGA in broader application scenarios.
\end{itemize}

\makefigpdf{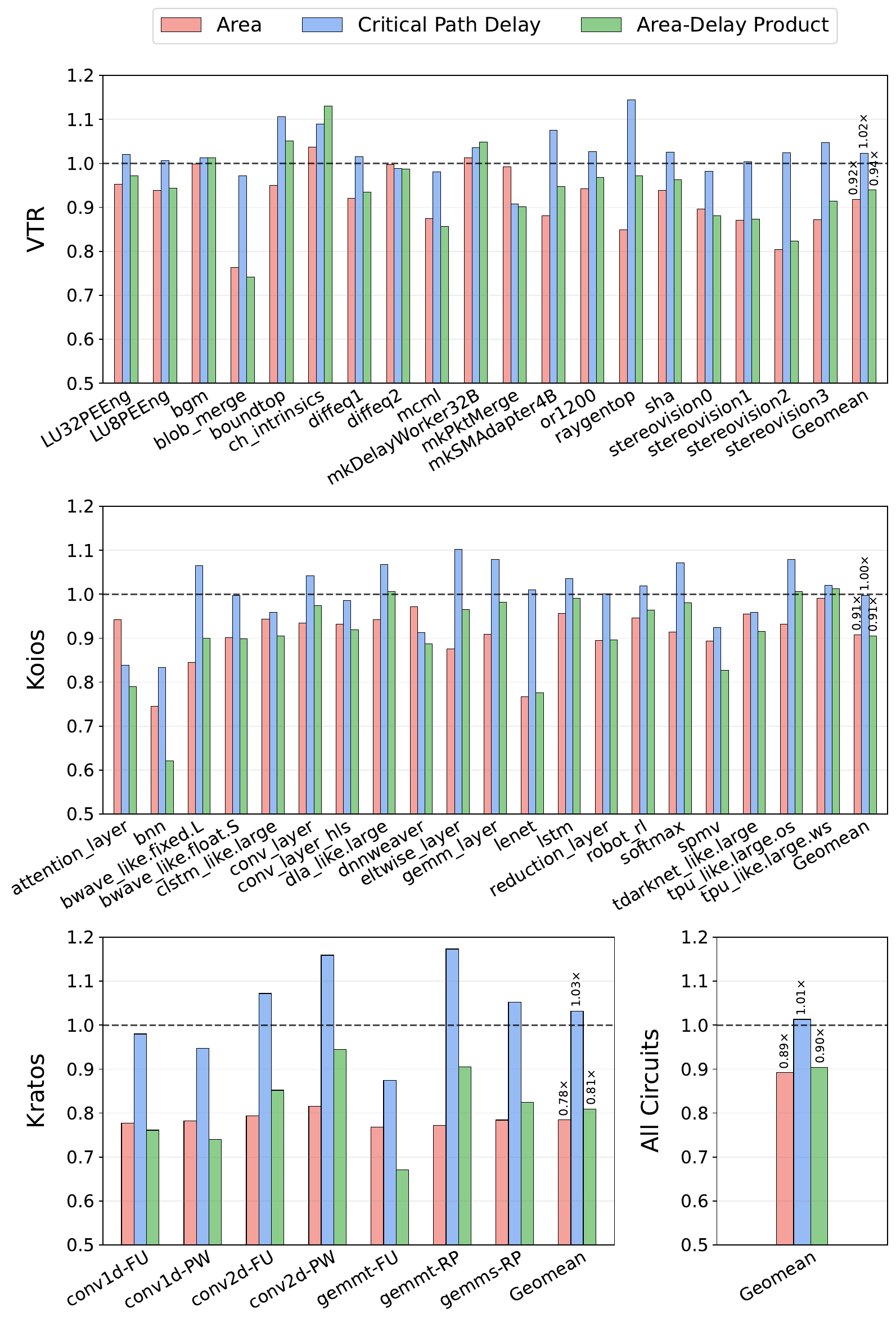}{Normalized ALM area usage, critical path delay, and area-delay product of \DD{} \ddfive Architecture, evaluated on the Koios, VTR, and Kratos benchmarks.}

\subsection{Experiment Results}

\textbf{Area and Delay Evaluation} 
\Cref{fig:results_base_vs_5lut} presents the normalized results of our proposed \ddfive architecture compared to the baseline Stratix 10 architecture across the Koios, VTR, and Kratos benchmark suites. In terms of ALM area usage, \ddfive consistently reduces resource consumption, achieving an average reduction of 10.9\%. This improvement is particularly substantial in the Kratos benchmark suite, where arithmetic operations dominate, showing an average of 21.6\% area savings.
While the average of critical path delay remains the same level as the baseline, a few circuits do exhibit an increase of up to 16\%. 
These cases are likely due to suboptimal packing, increased routing congestion, and routing heuristics that are not yet optimized for our modified architecture. 
Lastly, when considering the area-delay product (ADP), \ddfive demonstrates an average improvement of 9.7\%, indicating the effectiveness of our new \DD Architecture.

\makefigpdf{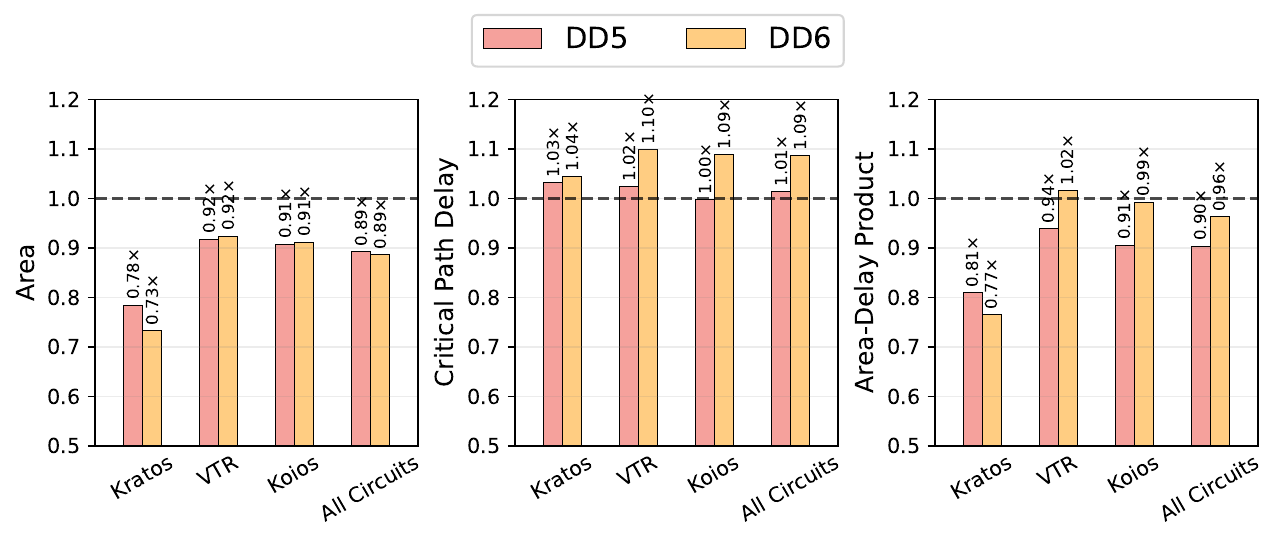}{Normalized ALM area used, critical path delay, and area-delay product (ADP) comparison between \ddfive and \ddsix architectures on the geometric means of 13 VTR standard benchmarks, 18 Koios benchmarks and 7 Kratos benchmarks set at data width of 6 and sparsity of 50\%.} 

\textbf{DD5 vs. DD6} We additionally evaluate the second variant \ddsix, which allows concurrent and independent usage of 6-LUT and adders, and should bring more flexibility compared to \ddfive. 
\Cref{fig:5lut_vs_6lut} shows the comparison between \ddfive and \ddsix. 
While \ddsix provides minor additional area savings in Kratos benchmarks, no noticeable gains are observed in Koios or VTR benchmarks. 
This can be attributed to two factors: (1) the current CAD tools are not optimized for architectures supporting concurrent 6-LUT usage, and (2) the utilization of 6-LUTs is inherently low.
On average only 7\% of ALMs across these benchmarks make use of 6-LUTs, making the opportunity for concurrent usage limited. 
Additionally, the added complexity of the output multiplexers in \ddsix leads to higher critical path delays, with an average frequency penalty of approximately 8\%.
Consequently, the area-delay product also increases, suggesting that the added flexibility of \ddsix does not offer practical benefits for general workloads.

\figvs{0.75}{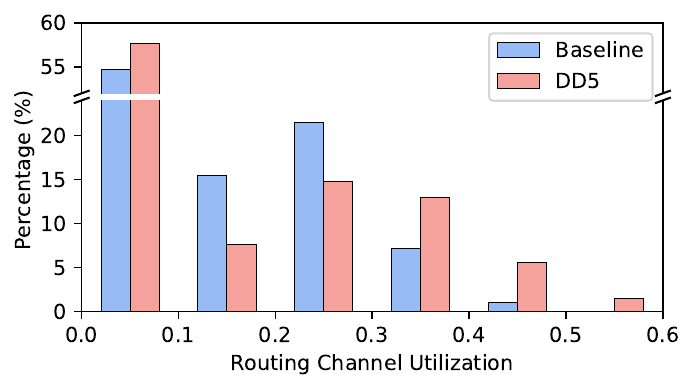}{}{Histogram of routing channel utilization averaged over the Kratos benchmark set. DD5 increases routing congestion on average as the diagram shows a shift of the histogram to higher channel utilization ranges.}

\textbf{Routing Congestion}
To analyze the routing impact of the proposed DD5 architecture, we examined the distribution of routing channel utilization using the Kratos benchmark set, which showed the largest area saving (22\%), and the highest critical path delay penalty (3\%) on average. 
The result is shown in \Cref{rcu_kratos}. 
While the portion of underutilized channels slightly increased, the majority of the channel utilization shifted towards a higher range, especially the portion between 0.3 and 0.6. 
This pattern suggests that by packing the same amount of logic into fewer logic blocks (LBs), routing channels become more congested, leading to denser usage.
For all the circuits in these benchmarks, none failed to route, and the critical path delay impact is not too high, suggesting that this additional routing congestion is well within the range of what the FPGA's routing network can handle.

\makefigpdf{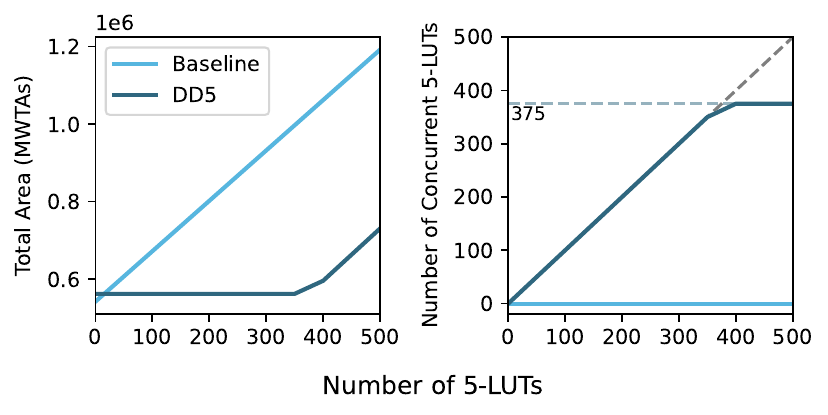}{Results of Packing Stress test. It starts with a circuit of 500 adders, and we pack an increasing number of LUTs into the circuit. The x-axis is the number of LUTs packed which saturates at 375 concurrent LUTs.}

\begin{table*}[t]
\centering
\caption{Results of the End-to-End stress tests. The table compares the maximum number of additional SHA instances packed into a fixed-size FPGA for three Kratos circuits using both the \base and the \ddfive architecture. The \ddfive architecture enables significant improvements in packing density and slight reductions in critical path delay.}
\label{tab:sha-stress}
\resizebox{\textwidth}{!}{
\begin{tabular}{rbacbacbac}
\toprule
\textbf{Base Kratos Circuit}        & \multicolumn{3}{c}{\bf conv1d-FU-mini}       & \multicolumn{3}{c}{\bf conv2d-FU-mini}    & \multicolumn{3}{c}{\bf gemmt-FU-mini}           \\
                                    \cmidrule(l){2-4}                              \cmidrule(l){5-7}                          \cmidrule(l){8-10}
\textbf{Architecture}               & \bf Base & \bf DD5          & \bf $\Delta$\%   & \bf Base  & \bf DD5      & \bf $\Delta$\%   & \bf Base & \bf DD5          & \bf $\Delta$\%      \\
\midrule
Maximum SHA instances               & 5        & 9                & +80.0\%        & 3        & 5             & +66.7\%         & 11       & 13              & +18.2\%           \\
\midrule
Adders ($\times10^3$)               & 37.55    & 38.79            & +3.29\%        & 18.86    & 19.48         & +3.28\%        & 14.35    & 14.97            & +4.31\%           \\
5-LUTs ($\times10^3$)               & 9.21     & 15.90            & +72.5\%        & 5.30     & 8.82          & +66.5\%        & 19.74    & 23.54            & +19.2\%           \\
Concurrent 5-LUTs                   & 0        & 4397             & +27.7\%$^*$        & 0        & 2458          & +27.9\%$^*$        & 0        & 3790             & +16.1\%$^*$           \\
\midrule
Critical Path Delay (ns)            & 15.81    & 15.05            & -4.81\%        & 16.07    & 15.01         & -6.60\%        & 16.27    & 15.16            & -6.82\%           \\
ALM Count ($\times10^3$)            & 27.93    & 27.08            & -3.04\%        & 14.16    & 13.79         & -2.64\%        & 18.45    & 18.04            & -2.27\%           \\
Logic Block Count                   & 2947     & 2946             & -0.03\%        & 1474     & 1498          & +1.63\%        & 1895     & 1870             & -1.32\%           \\
Total ALM Area (MWTAs $\times10^6$) & 60.54    & 60.88            & +0.57\%        & 30.69    & 30.99         & +0.98\%        & 40.00    & 40.55            & +1.37\%           \\
\bottomrule
\multicolumn{10}{c}{$^*$\scriptsize{Denotes the percentage of 5-LUTs in which LUTs and adders are used concurrently. This is impossible in the baseline architecture.}}
\end{tabular}
}
\end{table*}

\textbf{Packing Stress Test}
To quantify the limits of concurrent 5-LUT usage in \ddfive, we performed an \textit{artificial} packing stress test.
In this test, we constructed a synthetic circuit with 500 adders and incrementally added 5-LUT logic to the circuit, up to 500 LUTs in total. Under perfect concurrency conditions, this would be the maximum number of LUTs that could be absorbed into ALMs, concurrently with adders, without increasing utilization. 
We enabled the ``allow unrelated clustering'' option in VPR to encourage maximum packing density at the cost of ignoring timing, hence the ``artificial" nature of this test.
\Cref{fig:artificial_stress_test_results} shows the results. 
The left plot presents total area usage in minimum-width transistor areas (MWTAs). 
When only adders are present, the \base architecture has a slight area advantage since the \ddfive architecture has a small area overhead in each ALM. 
However, when LUTs are added, the \ddfive architecture shows a significant advantage as it can absorb LUTs into existing ALMs with adders, keeping the area the same until the ALMs are saturated. 
The right plot shows the number of concurrently packed 5-LUTs, which saturates at 375, corresponding to 75\% of the theoretical maximum. 
This indicates that up to 75\% concurrency between LUTs and adders is achievable, before becoming constrained by other factors like the global and local interconnect.

\textbf{End-to-End Stress Test}
In this \textit{realistic} stress test, we aim to evaluate how effectively the \DD architecture improves packing density under constrained FPGA resources. 
The procedure is as follows: we implement a circuit from the Kratos benchmarks to determine the FPGA size needed for a successful implementation. 
Then, keeping this FPGA size fixed, we incrementally add instances of a small SHA circuit from the VTR standard benchmark suite, until VTR can no longer complete placement and routing.
This is indeed how new FPGA architectures are stress-tested in industry~\cite{ilya}.
We conducted this experiment using both the \base architecture and the \ddfive architecture on three different Kratos circuits, and all experiments are performed with timing-driven placement and routing to better reflect a realistic scenario.  
\Cref{tab:sha-stress} shows the results, indicating that \ddfive enables significant improvements in packing density. 
For instance, the maximum number of SHA instances that can be implemented in the same FPGA area increased by 80\% for \texttt{conv1d-FU-mini} and by 66.7\% for \texttt{conv2d-FU-mini}, compared to the \base. 
This is achieved through significant 5-LUT concurrency: 27.7\% for \texttt{conv1d-FU-mini}, 27.8\% for \texttt{conv2d-FU-mini}, and 16.1\% for \texttt{gemmt-FU-mini}. 
Additionally, the critical path delay also shows slight improvements; this is because the critical path lies in the multiple adder region, and our AddMux in \ddfive architecture reduced the delay from ALM input to the adder, resulting in a lower overall critical path delay. 
These results demonstrate that \ddfive not only allows for denser packing but can also improve timing in certain stress scenarios.

\section{Discussion}

\textbf{Related Works}
There are two closely-related works targeting the arithmetic efficiency of the FPGA's soft logic. 
Eldafrawy et al.~\cite{eldafrawy:acm2020:s10-mod} proposed modifications to the ALM microarchitecture with an additional adder chain. 
However, their work's main focus is on low-bit multiply-accumulate (MAC) only, and lacks end-to-end evaluations on circuit benchmarks, making it challenging to compare directly to their work.
Furthermore, our approach is synergistic with this prior work as we can bypass the LUTs to access one or more adder chains within the ALM.
Similarly, LUXOR ~\cite{rasoulinezhad:fpga20:luxor} achieves higher arithmetic efficiency of compressor trees by improving the expressiveness of each ALM through an extra XOR gate. 
This approach is also orthogonal to ours, as LUXOR focuses on gate-level function enhancement, whereas \DD focuses on improving data path flexibility and resource utilization. 
Therefore, both enhancements can be applied to the same FPGA simultaneously.

\textbf{Compatibility with Other Architectures}
Our architectural exploration and evaluation are based on an Intel Stratix-10-like FPGA, as its open-source architectural models are readily available~\cite{eldafrawy:acm2020:s10-mod}. However, modern FPGA families—such as the AMD Versal series~\cite{amdlut}—feature different design paradigms: they use LUTs as full adders and include dedicated carry-lookahead logic, fundamentally different from the Stratix 10. 
To adapt the \DD architecture to these FPGAs, several function unit augmentations would be necessary, including reintroducing explicit 1-bit full adders. While this requires some additional logic, the relative area cost of full adders in modern technologies is negligible and unlikely to cause substantial overhead. As a result, we believe our \DD architecture is adaptable to modern devices with minimal design changes and without compromising its core efficiency benefits.
We plan to more accurately compare to these alternative logic block architectures in future work.

\section{Conclusion}

We presented \DD, a novel FPGA logic block architecture that enables the concurrent and independent usage of adders and LUTs within the same ALM, significantly improving arithmetic density with minimal area overhead and a negligible impact on critical path delay. 
Together with the architectural innovations, we enhanced the VTR CAD toolchain\cite{vtr9} with optimized synthesis algorithms for adder chains, substantially improving its efficiency and providing us with a baseline that is representative of commercial CAD tools like Quartus. 
We evaluated \DD across three diverse benchmark suites,
achieving an average 9.7\%  improvement in area-delay product and up to 21.6\% area savings on the arithmetic-heavy Kratos benchmarks~\cite{dai:fpl24:kratos}. 
Although denser packing may introduce routing pressure, we found the increase in routing congestion to be relatively low and well within the capabilities of modern FPGA routing fabrics. 
Lastly, two stress tests show the potential of \DD in achieving higher density and resource utilization.

\section*{Acknowledgment}
{\small
This project is supported by Intel Corporation, the National Science Foundation under Grant No. 2303626, and the CN Yang Scholars Programme, Nanyang Technological University.
We would like to thank Sergey Gribok, Ilya Ganusov, Martin Langhammer, Sadegh Yazdanshenas, and Susanne Balle for discussion and feedback.
}

\bibliographystyle{IEEEtran}
\bibliography{IEEEabrv,main,confs}
\end{document}